\documentclass[column,secnumarabic,amssymb,amsmath,nofootinbib,floatfix,
nobibnotes,aps,
preprint,
superscriptaddress,
prd
]{revtex4}

\usepackage[dvipsnames]{xcolor}
\usepackage{caption}
\usepackage{amsmath,adjustbox,mathtools,amssymb}
\usepackage{amsfonts}
\usepackage{amssymb}
\usepackage{graphicx}
\usepackage{braket}
\usepackage{slashed}
\usepackage{wrapfig}
\usepackage{tikz-feynman,contour}
\usepackage{tikz-feynhand}
\usepackage{feynmp-auto}
\usepackage{bbold}
\usepackage{bm}
\usepackage{verbatim}
\usepackage{subcaption}
\usepackage{appendix}
\usepackage{slashed}
\usepackage{ulem}
\normalsize
\usepackage{booktabs}
\usepackage{supertabular}
\usepackage{tabularx}
\usepackage{tabulary}
\usepackage{multirow}
\usepackage{soul}

\newcommand{\ud}{{\rm{d}}}

\begin{document}

\title{MEC-induced two-nucleon emission in neutrino-nucleus scattering}


\author{M.B. Barbaro} 
\affiliation{
	Dipartimento di Fisica, Universit\`a di Torino, via P. Giuria 1, 10125 Turin, Italy}
\affiliation{
	INFN, Sezione di Torino, via P. Giuria 1,10125, Turin, Italy
}
\author{V. Belocchi} 
\affiliation{
	Dipartimento di Fisica, Universit\`a di Torino, via P. Giuria 1, 10125 Turin, Italy}
\affiliation{
	INFN, Sezione di Torino, via P. Giuria 1,10125, Turin, Italy
}
\affiliation{
Instituto de Física Corpuscular (IFIC), Consejo Superior de Investigaciones Científicas (CSIC) and Universidad de Valencia, E-46980 Paterna, Valencia, Spain
}
\author{A. De Pace} 
\affiliation{
	INFN, Sezione di Torino, via P. Giuria 1,10125, Turin, Italy
}
\author{M. Martini}
\affiliation{IPSA-DRII,  63 boulevard de Brandebourg, 94200 Ivry-sur-Seine, France}
\affiliation{Sorbonne Universit\'e, CNRS/IN2P3, Laboratoire
de Physique Nucl\'eaire et de Hautes Energies (LPNHE), Paris, France}

\begin{abstract}

Meson-exchange currents (MEC) play a crucial role in the nuclear response to a leptonic probe, inducing the emission of two nucleons, the so-called 2p2h  excitations. 
Here we report the results of our recent work \cite{Belocchi:2025eix}, in which we evaluated the 2p2h  contribution to the  $(\nu_\mu,\mu^-p)$  neutrino cross section off a carbon target,  which involves the simultaneous detection the outgoing muon and a proton in the final state. This process has recently been measured in long-baseline neutrino experiments, but  no truly microscopic calculation for the 2p2h  channel existed to be compared with these data until now.  

The calculation employs fully relativistic two-body currents and is carried out within the Relativistic Fermi Gas framework. It
 provides a generalization to the weak sector of the electromagnetic inclusive model developed in Ref.~\cite{DePace:2003spn}  and an extension of the calculation to enable semi-inclusive predictions. A selection of our results for the semi-inclusive cross section versus the emitted proton variables is presented, both at fixed neutrino energy and folded with the T2K muon-neutrino flux. 

\end{abstract}

\maketitle

\section{Introduction}

Understanding neutrino properties — especially the PMNS parameters and CP-violating phase — is a key goal of long-baseline experiments like T2K, NOvA, DUNE and Hyper-Kamiokande. Precise modeling of neutrino-nucleus interactions is essential for these experiments, as nuclear targets are required due to neutrinos' weak interaction with matter~\cite{NuSTEC:2017hzk}.

A major challenge in long-baseline experiments is reconstructing the incoming neutrino energy $E_\nu$, which must be inferred from the final state due to the wide energy distribution of the flux. This requires consistent modeling of all interaction channels — in particular the complex two-particle–two-hole (2p2h) mechanism, involving two-nucleon emission via two-body currents.

The importance of 2p2h  was first recognized when the first MiniBooNE data of neutrino-carbon cross sections without pions in the final state were published~\cite{MiniBooNE:2013qnd},  showing discrepancies with Fermi gas model predictions. Theoretical work incorporating 2p2h~\cite{Martini:2009uj,Nieves:2011pp,Megias:2016fjk} helped explain this discrepancy and this channel is now widely accepted as a crucial contribution to the inclusive $\nu_l,l)$ cross section.
Multiple models exist for this process, but theoretical uncertainties remain significant.

Semi-inclusive measurements — where hadrons are detected in coincidence with the lepton — offer tighter constraints on nuclear models. These observables are more sensitive to nuclear dynamics and have driven interest in models that go beyond inclusive-only approaches. While the quasi-elastic (QE) semi-inclusive channel has seen some recent progress — see for example the calculation of Ref.~\cite{Franco-Patino:2020ewa} — a fully microscopic treatment of semi-inclusive 2p2h  process is still missing. Despite this, 2p2h  effects significantly influence semi-inclusive cross sections, and their improper treatment affects conclusions about nuclear models and neutrino data interpretation. 

While several semi-inclusive data have now become available, current event generators used in experimental analyses simulate 2p2h  contributions using inclusive inputs, which are inadequate for predicting semi-inclusive or exclusive observables.
To address this gap, we have developed a fully microscopic model for semi-inclusive 2p2h  neutrino-nucleus interactions, extending  the electromagnetic 
$(e,e'p)$ framework of Ref.~\cite{Belocchi:2024rfp} to the weak sector, and so enabling realistic implementation in event generators.

\section{Formalism and model}
\label{sec:Formalism}

Let us briefly outline the basic formalism required  to describe the charged-current (CC) semi-inclusive neutrino-nucleus interactions. Further details can be found in Refs.~\cite{Belocchi:2025eix,Belocchi:2025lqp}.
We focus on the process  in which a muon neutrino of energy $E_k$ scatters off a nucleus $A$ at rest in the laboratory frame via the charged current process:
\begin{equation}
\nu_\mu+A \to \mu^-+N+X \,.
\label{eq:eepN}
\end{equation}
  The final state consists of a muon \(\mu^-\),  with momentum $k'$ and solid angle $\Omega_{k'}$, a knocked-out nucleon, with momentum $p_N$, energy $E_N$ and solid angle $\Omega_{N}$, and the residual system $X$. 
The six-fold differential semi-inclusive cross section with respect to the muon and nucleon variables is given by
\begin{equation}
    \frac{\rm d^6 \sigma}{\rm d E_{k'}\rm d \Omega_{k'} \ud E_{N} \ud \Omega_{N}}=
    \frac{G_F^2}{8\pi^2}\cos^2\theta_C\frac{k'}{E_k}
 p_N E_N \,\widetilde L_{\mu \nu} W^{\mu \nu}_{A(N)}\,,
    \label{eq:F2def}
    \end{equation}
which involves a contraction of the weak lepton tensor $\widetilde L_{\mu\nu}$
with the semi-inclusive weak nuclear tensor \cite{Belocchi:2024rfp}
\begin{equation}
    W^{\mu\nu}_{A(N)} = \sum_X \langle A|\hat J^{\mu\dagger}|N,X\rangle \langle N,X| \hat J^\nu | A\rangle
    \,\delta\left(E_{N}+E_X-E_0-\omega\right) \,.
    \label{eq:WmunuAN}
\end{equation}
This tensor encapsulates all nuclear effects, including initial- and final-state interactions, nuclear recoil, and the energy absorbed by the residual system, thus reflecting all the features of the adopted nuclear model.

In Eq.~\eqref{eq:WmunuAN}, 
$|A\rangle$ denotes the nuclear ground state of energy $E_0$, the ket $|N,X\rangle$ is the hadronic final state of energy $E_N+E_X$, and a sum over the unobserved states $X$ is performed. 
The weak nuclear current is given by
\begin{equation}
\hat J^\mu =  \hat J^\mu_{1b} +\hat J^\mu_{2b}+...+\hat J^\mu_{Ab} \,,
\end{equation} 
where $\hat J^\mu_{nb}$ denotes the $n$-body current, with $n=1,\,\cdots A$.
Consequently, the hadronic tensor  can be expressed as the sum of various contributions corresponding to the excitation of different final states. Here we retain only the one- and two-body currents, which can excite one-particle-one-hole (\(X=\rm 1p1h\)) and two-particle-two-hole (\(X=\rm 2p2h \)) states:
\begin{equation}
    W^{\mu\nu}_{A(N)} \simeq W^{\mu\nu}_{A(N), \rm 1p1h}+W^{\mu\nu}_{A(N),\rm 2p2h }\,,
    \label{eq:W1+W2}
\end{equation}
and we focus in particular on $W^{\mu\nu}_{A(N),\rm 2p2h }$. Note, however, that the two-body current contributes also to the 1p1h tensor through interference with the one-body current~\cite{Casale:2025wsg}.

Compared to the electromagnetic case studied in Ref.~\cite{Belocchi:2024rfp},
 the weak axial current introduces additional  contributions.  The non-conservation of the axial current, as well as vector–axial interference, significantly complicates neutrino CC semi-inclusive scattering relative to electron scattering case. Furthermore, the lack of azimuthal symmetry in the semi-inclusive case gives rise to ten non-vanishing response functions, as opposed to five in the inclusive case (see Ref.\cite{Belocchi:2024rfp}).
 
To compute the 2p2h  hadronic tensor, we adopt as wave-function basis the Relativistic Fermi Gas (RFG) model, which treats the nucleus as a system of non-interacting nucleons described by Dirac spinors, correlated only by the Pauli exclusion principle.
The RFG model may seem inadequate to provide semi-inclusive predictions. This is certainly true in the quasi-elastic channel, where the RFG predicts that for a given nucleon observed in the final state, only one initial nucleon configuration is allowed, which is far from being realistic. However,  this limitation is mitigated in the 2p2h sector. In this case, the interaction involves two-body currents acting on distinct nucleons in the Fermi sea, introducing dynamical correlations that go beyond the simple Fermi gas picture.

The RFG-based 2p2h  nuclear tensor takes the form
\begin{equation}
\begin{aligned}
 & W_{A(N)\,2p2h }^{\mu \nu}= \frac{m_N^2}{(2\pi)^3E_{p_1}}
  \theta(|\mathbf{p_1}|-p_F|)
\left[ \prod_{i=1}^2 \int_{|\mathbf{h}_i|\leq p_F} \frac{m_N \, \ud \mathbf{h_i}}{(2\pi)^3E_{h_i} }\right]
\\  &\times \int_{|\mathbf{p_2}|\geq p_F} \frac{\ud \mathbf{p_2}}{E_{p_2}}  \, w_{\rm 2p2h }^{\mu \nu}(h_1,h_2,p_1,p_2)\, \delta^4(q +h_1+h_2 -p_1-p_2)\,,
  \end{aligned} 
  \label{eq:WmunuN}
  \end{equation} 
  where the nucleon detected has been arbitrarily chosen to be the one with momentum \(\mathbf{p_1}\), without loss of generality. 
  Pauli blocking is encoded in the two step functions acting on the final-particle momenta $\mathbf{p_1}$ and $\mathbf{p_2}$.
   
The elementary tensor appearing in Eq.~\eqref{eq:WmunuN} is defined as follows:
\begin{equation}
     w^{\mu\nu}_{2p2h }(h_1,h_2,p_1,p_2) = \frac{1}{4}      \sum_{\substack{\mathrm{spin}\\ 
    \mathrm{isospin*}}}\langle h_1 h_2|\hat J^{\mu\dagger}_{2b}|p_1p_2\rangle \langle p_1p_2 | \hat J^\nu_{2b} | h_1h_2\rangle ,
    \label{eq:reduced_w}  
\end{equation}
where $\hat J^\mu_{2b}$ is the two-body current operator, while the kets \( \ket{h_1 h_2} \) and \( \ket{p_1 p_2} \) indicate the initial and final two nucleon states, respectively, characterized by their momenta.
The sum runs over all isospin configurations that include at least one nucleon of the detected type (proton or neutron), but excludes the final-state isospin of the detected particle.

\begin{figure}[ht!]
   \centering
\includegraphics[width=.7\linewidth]{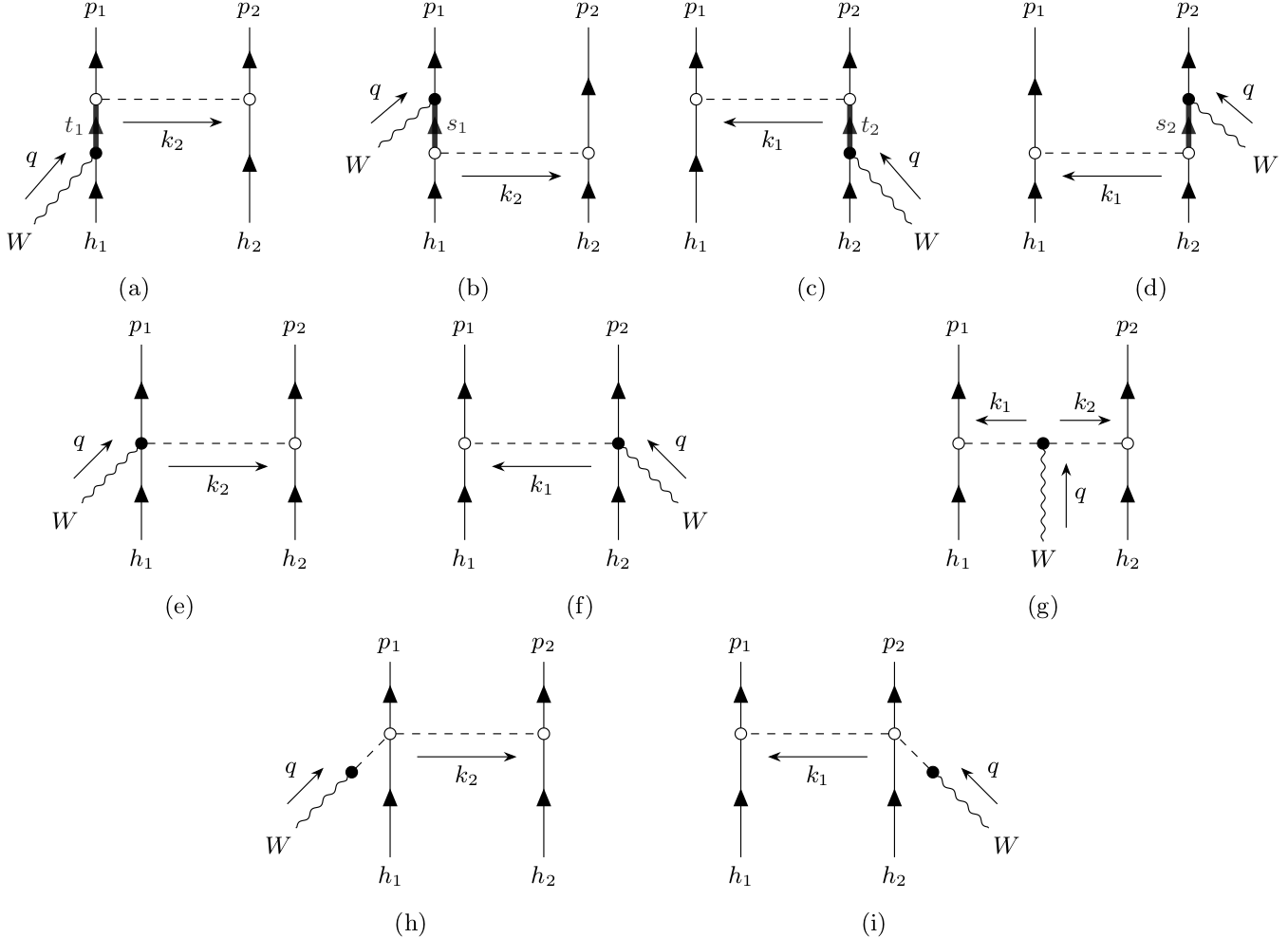}
\caption{Feynman diagrams of the weak CC MEC current, involving nucleons (solid lines), the $\Delta$ resonance (thick lines) and the pion (dashed lines) .}
    \label{fig:2p2h } 
\end{figure}

The two-body current operator we adopt to evaluate the tensor \eqref{eq:reduced_w} is a Meson Exchange Current (MEC),
obtained from a chiral Lagrangian which describes the interaction between nucleons, pions and the $\Delta$ resonance \cite{Hernandez:2007qq}.
It consists of five components:
\begin{equation}
\hat J^\mu_\mathrm{MEC}\equiv \hat J^\mu_\mathrm{pif}+\hat J^\mu_\mathrm{sea}+\hat J^\mu_\mathrm{pp}+\hat J^\mu_{\Delta_\mathrm{F}}+\hat J^\mu_{\Delta_\mathrm{B}}\,,
\end{equation}
denoted as pion-in-flight (pif), seagull (sea),  pion-pole (pp) and the $\Delta$-resonance - forward (F) and bacward (B) -  and represented by the  Feynman diagrams shown in Fig.~\ref{fig:2p2h }. Explicit expressions of these currents are given in \cite{Belocchi:2025eix}.

To make the RFG model more realistic, an energy shift is introduced, which phenomenologically accounts
for the nucleon binding energy and for final-state interaction effects. 
This energy shift, denoted $E_{\rm shift}$,  is treated as a constant parameter and extracted phenomenologically from electron–nucleus scattering data by matching the position of the quasi-elastic peak.
In practice, this means that part of the transferred energy $\omega$ is used to overcome nuclear binding. The elementary tensor is thus evaluated using an effective energy transfer \(\tilde \omega = \omega-E_{\rm shift}\).
Following the de Forest prescription~\cite{DeForest:1983ahx}, we use the true $\omega$ for evaluating the form factors appearing in the current operators, while the rest of the hadronic tensor is modified by the energy shift.  For the carbon nucleus considered in this work, on the basis of  the study \cite{Maieron:2001it} of $(e,e')$ data and of our previous analysis of $^{12}$C$(e,e'p)$ reactions~\cite{Belocchi:2024rfp}, we adopt  the vale \(E^{\rm 2p2h }_{\rm shift}= 2 E_{\rm shift} = 40\) MeV. For the other parameter of the RFG, the Fermi momentum, fitted to the width of the QE peak, we use $p_F=225$ MeV.

The evaluation of the 2p2h semi-inclusive hadronic tensor \eqref{eq:WmunuN} involves a nine-dimensional integral. 
Through appropriate manipulations and by analytically exploiting the Dirac $\delta^4$, this integral can be reduced to a five-dimensional form, which is then computed numerically. 

\section{Results}
\label{sec:Results}

Before evaluating the semi-inclusive 2p2h cross sections, in Ref.~\cite{Belocchi:2025eix} we have validated our calculation by comparing our results for the inclusive $(\nu_\mu,\mu)$ responses with the ones published in Ref.~\cite{RuizSimo:2016ikw}, which employs the same model.  The comparison,  not shown here due to space limitations, reveals perfect agreement.

When considering a semi-inclusive process such as $(\nu_\mu,\mu N)$, the cross section is evaluated as a function of the three-momentum carried by one of the nucleons in the final state. Usually, this is chosen to be a proton, because it is the one that is easier to detect experimentally. Note that, unlike in the electromagnetic case, in CC neutrino-nucleus scattering a proton is always emitted in the 2p2h channel. 

To explore how the cross section depends on the final proton kinematics, we evaluated Eq.~\eqref{eq:F2def} by fixing the incident neutrino energy \(E_k\equiv E_{\nu_\mu}\), the muon scattering angle  \(\theta_{k'}\equiv\theta_\mu\) and the transferred energy \(\omega\). The reference frame considered is the q-system, so that each angle is relative to the \({\bf q}\) direction.

\begin{figure}[ht!]
    \centering
    \includegraphics[width=1\linewidth]{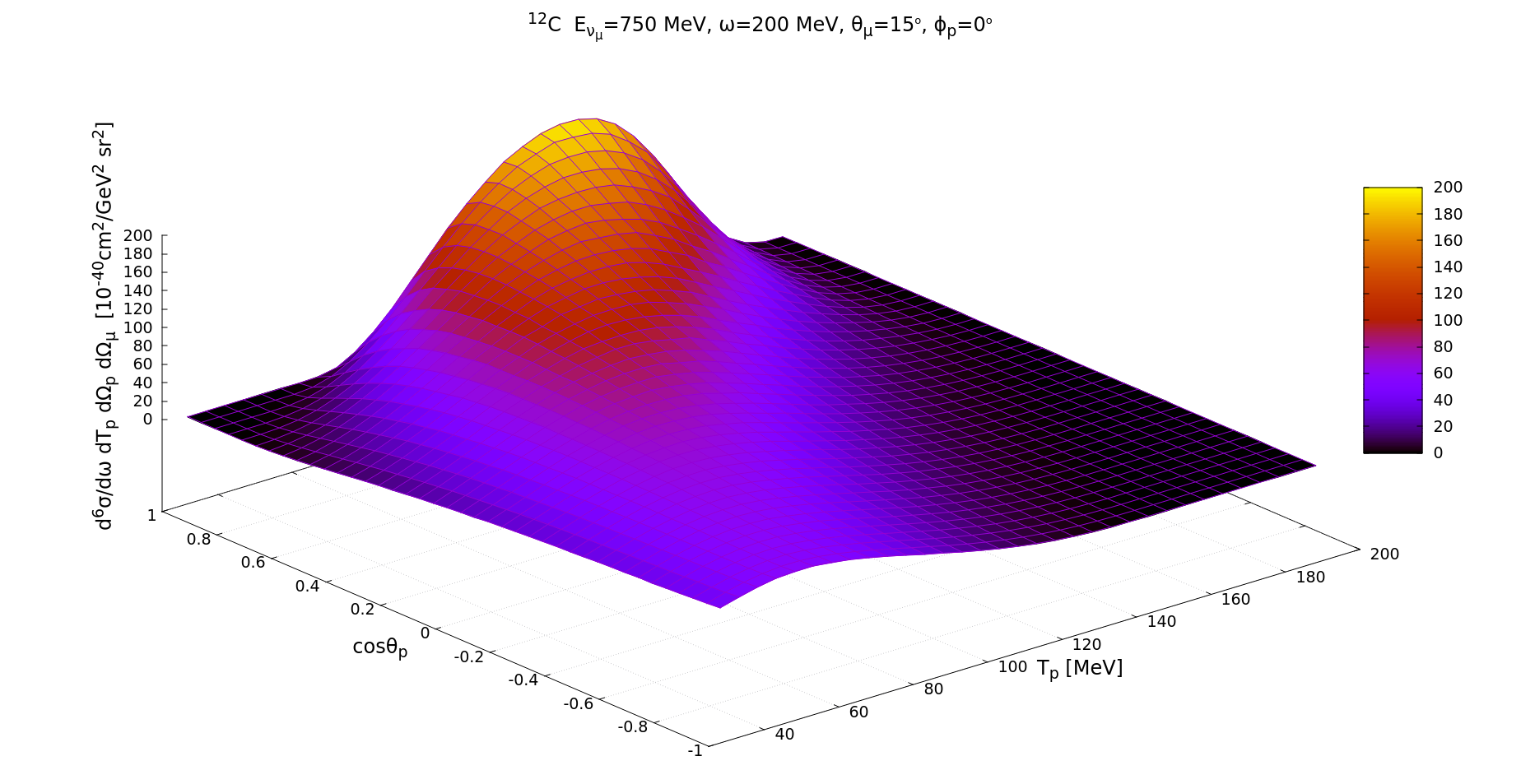}
    \caption{CC semi-inclusive $\nu_\mu$-$^{12}C$ sixth differential cross section for the 2p2h channel, computed at incident muonic neutrino energy \(E_{ \nu_\mu}=750\) MeV and transferred energy \(\omega=200\) MeV and displayed as a function of the polar angle \(\theta_p\) and kinetic energy \(T_p\) of the final proton. The scattering angle \(\theta_\mu=\)15° is fixed, as well the azimuthal final proton angle \(\phi_p=0\)°. }
    \label{fig:3D-EW15}
\end{figure}
        
In the three-dimensional plot of Fig.~\ref{fig:3D-EW15} the cross section is displayed for \(E_{\nu_\mu}=750\) MeV, \(\omega=200\) MeV and  \(\theta_\mu=15^\circ\) as a function of the proton polar angle and kinetic energy.
This kinematics is such that the 2p2h  channel gives an important contribution to the cross section: indeed in these conditions the cross section is dominated by 2p2h, the QE peak being ceneterd at \(\omega_{\rm QE}=Q^2/2m_N\simeq37\) MeV – including \(E_{\rm shift}\) – while the pion production, starting when the energy transfer is higher than the pion mass, is still small.
Under these conditions the 2p2h contribution exhibits a well-defined  peak  at \(\theta_p=0^\circ\), corresponding to the so-called ``parallel kinematics''. 
\begin{figure}[ht!]
    \centering
    \includegraphics[width=.45\linewidth]{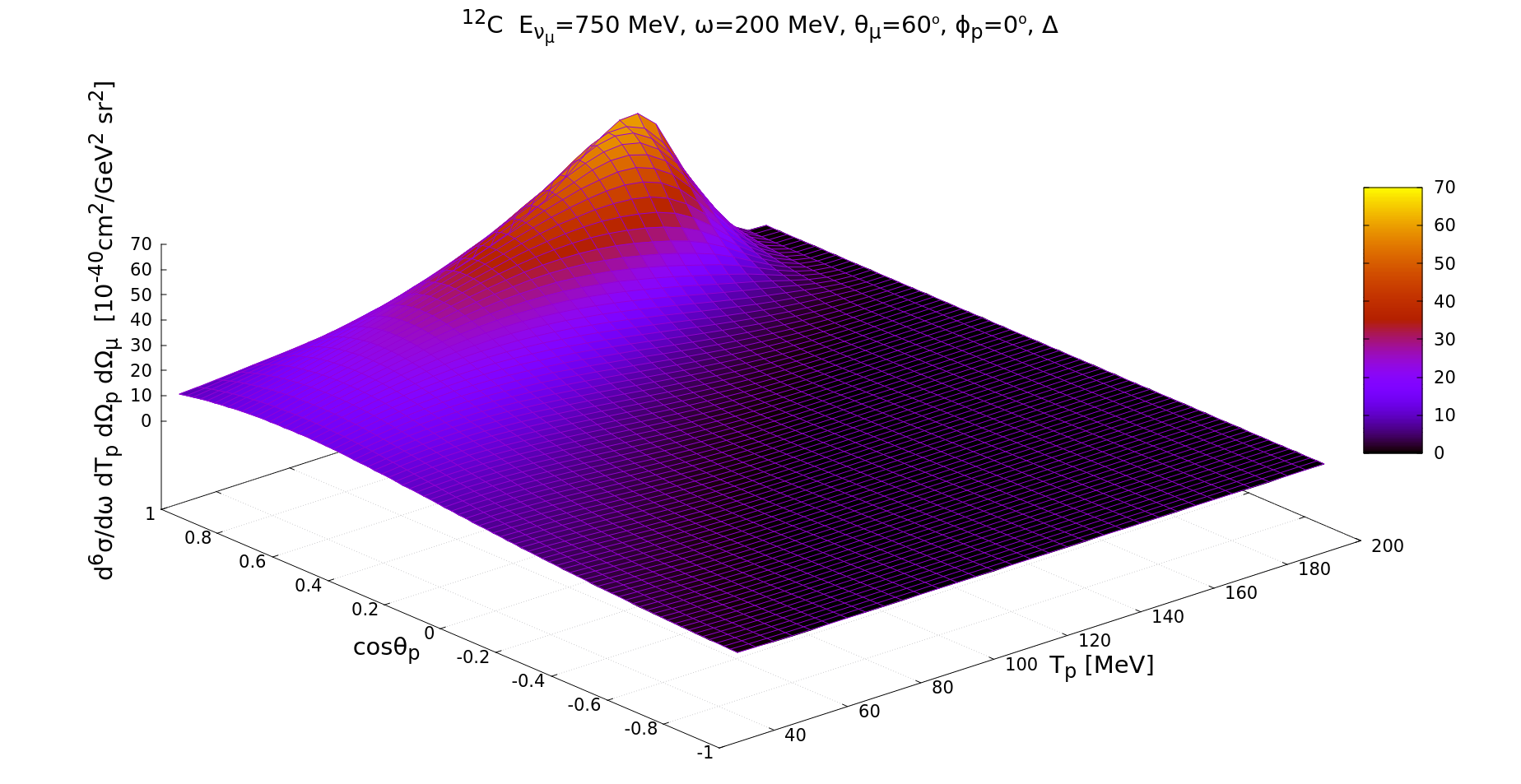} \includegraphics[width=.45\linewidth]{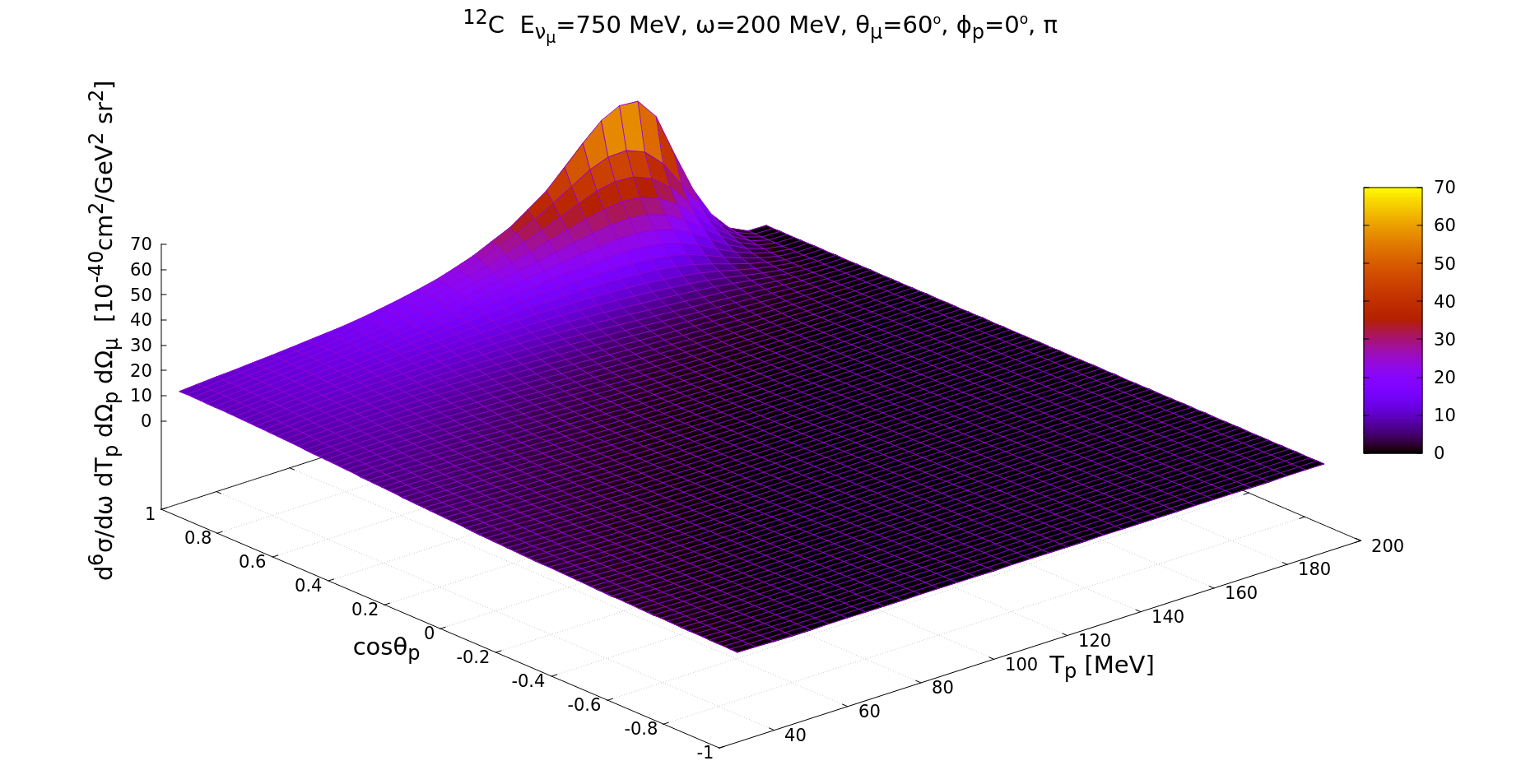}   \includegraphics[width=.45\linewidth]{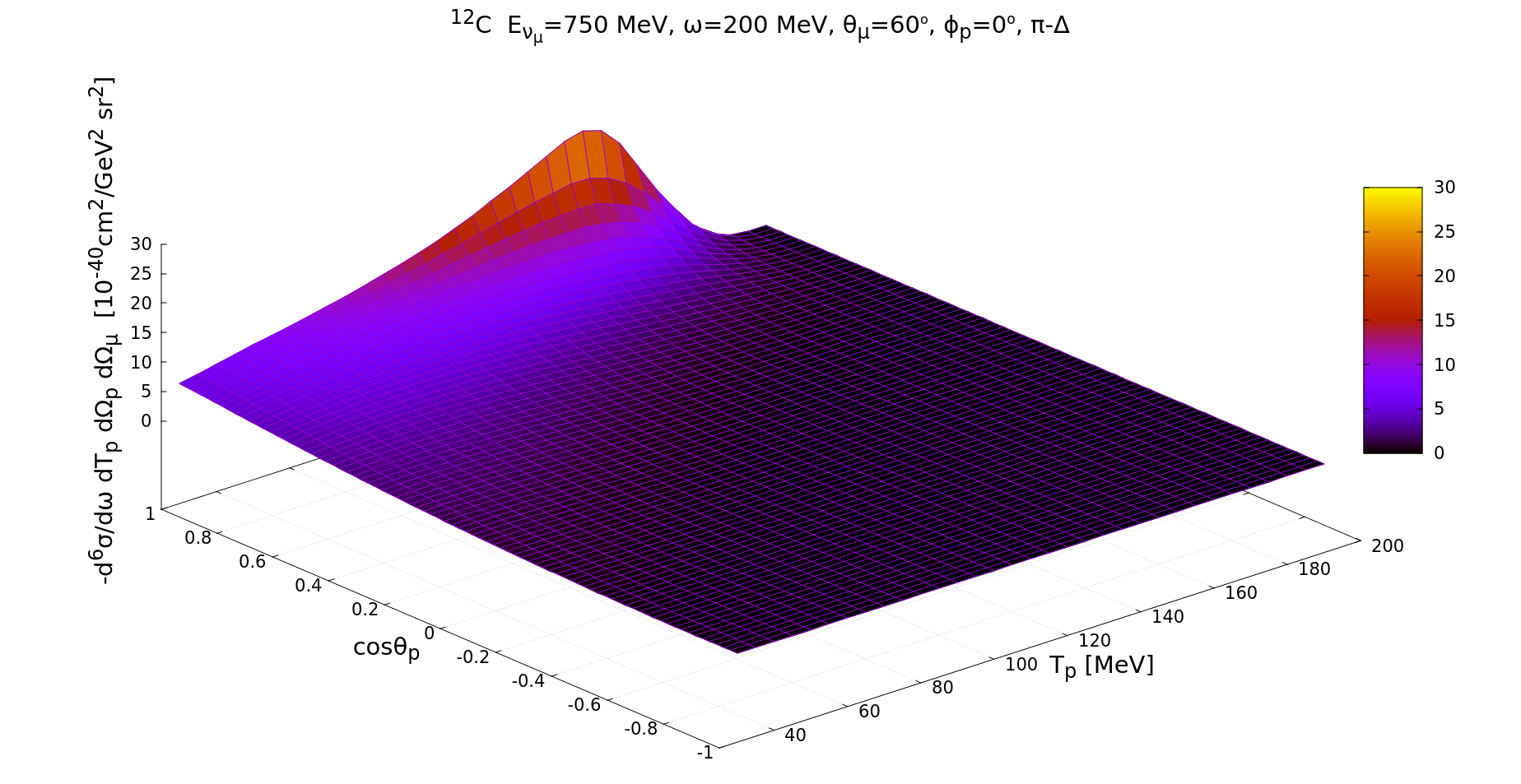} \\  
    \caption{Same as Fig.~\ref{fig:3D-EW15}, but for \(\theta_\mu = 60\)°  and with the \(\Delta\), pionic and \(\pi-\Delta\) interference displayed separately. Note the minus sign in the label definition for the last contribution.}
    \label{fig:3D-60deg_study}
\end{figure}

A detailed decomposition of the 2p2h contribution into meson-exchange current  components is illustrated in Fig.~\ref{fig:3D-60deg_study}, where the separate \(\Delta\), purely pionic and  interference \(\pi\)--\(\Delta\)  cross sections are displayed for  \(\theta_\mu=60^\circ\).
It clearly appears that the \(\Delta\) current dominates, accounting for more than half the total strength. This component exhibits a distinct peak around  \(T_p\simeq 150\) MeV
 in parallel kinematics, rapidly decreasing with increasing proton angle and kinetic energy. In contrast, the purely pionic current peaks at the same location but is more sharply localized, with a steeper drop-off away from the peak. The  \(\pi\)--\(\Delta\)  interference  is the smallest component, but still non-negligible and negative in this particular configuration. Both its sign and magnitude are highly sensitive to the specific kinematics. In this case, it displays a broad structure in proton kinetic energy, peaking under parallel kinematics and diminishing at larger angles, with a smoother tail around \(\theta_p\simeq 30\)°.

We next consider the single-differential cross section  \( \rm d \sigma/\ud p_N \), obtained by integrating Eq.~\eqref{eq:F2def} over the muon momentum and angle, and over the final proton angles. This quantity can also be expressed in terms of the momentum of the “leading” proton, defined as the most energetic proton in a multi-nucleon knockout event. In the 2p2h context, the leading proton corresponds to the more energetic nucleon in a  \(pp\)  final-state pair, or the only proton present in a  \(pn\)  final state.

\begin{figure}
    \centering
    \includegraphics[width=1\linewidth]{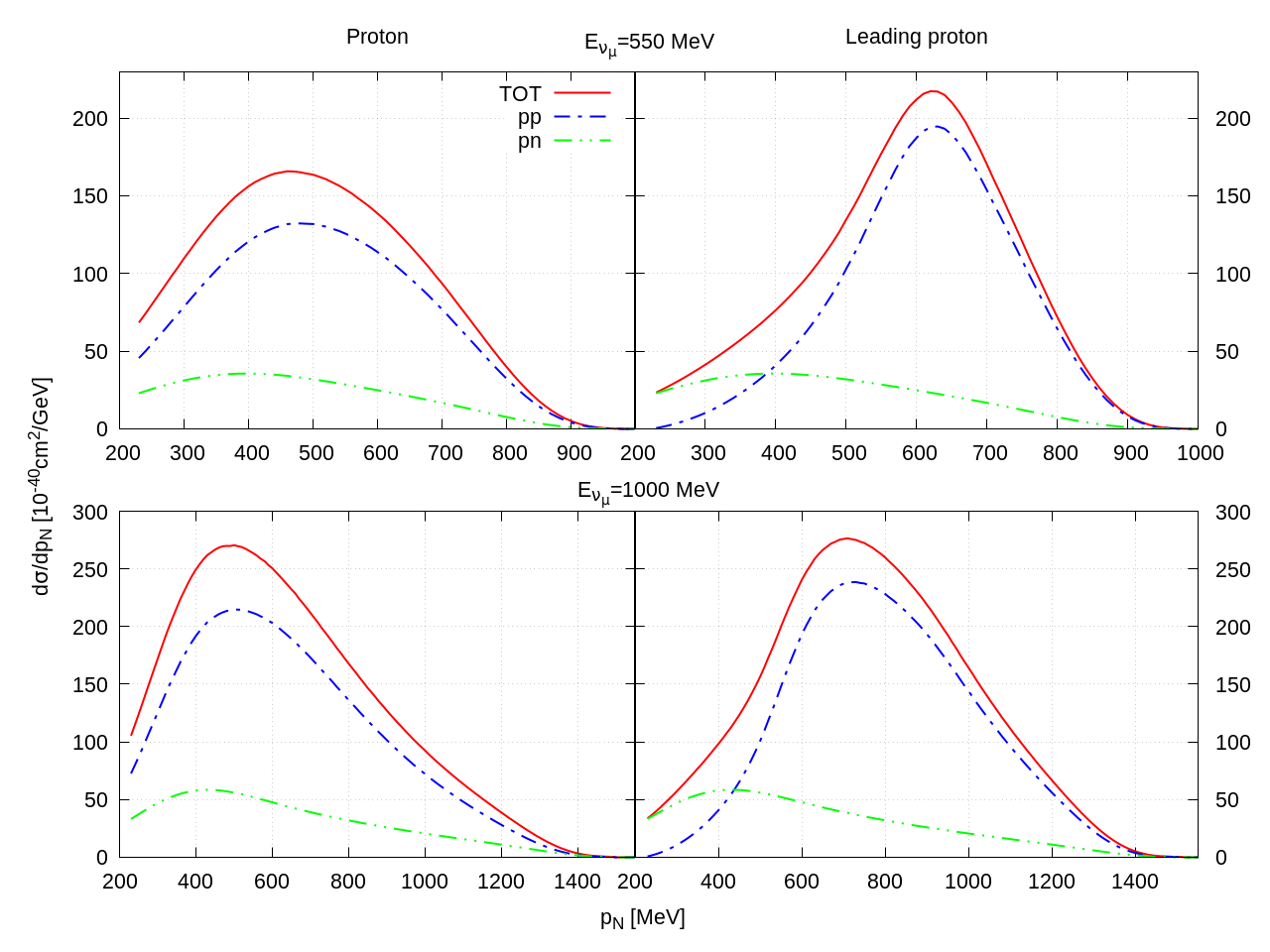}
    \caption{CC   semi-inclusive $\nu_\mu$-$^{12}C$ cross section evaluated at incident neutrino energy 550 (top) and 1000 (bottom) MeV. 
    In the left panels the cross section is computed varying the final proton momentum, while in the right ones the leading proton momentum is used.}
    \label{fig:EW-leadproton}
\end{figure}

The effect of the leading proton definition is examined in Fig.~\ref{fig:EW-leadproton} by comparing the differential cross section  \( \frac{\rm d \sigma}{\ud p_N} \), separated into its isospin channels, as a function of the proton momentum (left panels)  and of the leading proton momentum (right panels), for two fixed values of the neutrino energy. As shown, the contribution from the \(pn\) final state is identical in both approaches, while the \(pp\) final-state channel is significantly affected. This difference is expected: using the leading proton variable, when two protons are present in the final state, the computed cross section is determined by the most energetic proton. Consequently, the cross section associated with small final-proton momenta in the \(pp\) channel is significantly suppressed. 
This phenomenon is more pronounced at low incident neutrino energy, caused by the peak shape and position: it is broad and centered at lower final-proton momentum. In this region, the leading proton definition has a more significant impact, enhancing the peak, which shifts to higher proton momentum values, while reducing the tails. 
At higher neutrino energies, both representations yield similar distributions, though the leading proton variable still produces a modest shift of the peak to higher momenta.

Finally, to make contact with experimental data, we have calculated flux-folded  semi-inclusive cross sections, using the kinematics of the  T2K  $1\mu$CC$0\pi Np$ measurement \cite{T2K:2018rnz}, which corresponds to events with no pions and at least one proton in the final state. In the calculation we  have applied the experimental kinematic cuts, which impose significant constraints on the final-state phase space, including restrictions on the muon scattering angle and on both the polar angle and momentum of the outgoing proton in the laboratory frame.
\begin{figure}
    \centering
    \includegraphics[width=0.495\linewidth]{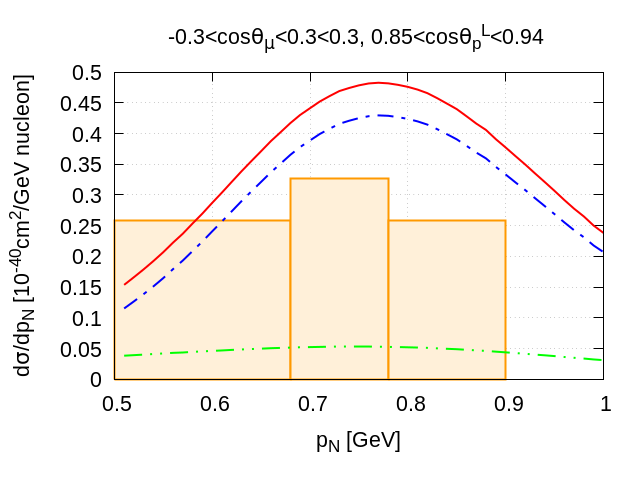} \includegraphics[width=0.495\linewidth]{ 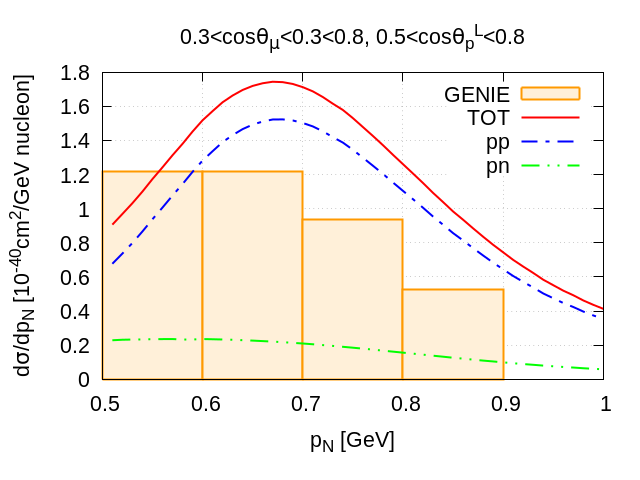}\\
    \caption{Differential  2p2h contribution to the \(1\mu\)CC\(0\pi Np\) cross section averaged over the incident T2K ND280 muon neutrino flux with T2K \cite{T2K:2018rnz} kinematical cuts compared to ``GENIE'' \cite{Dolan:2019bxf} theoretical predictions. The results are normalized for single active nucleon of  the mineral oil target CH.}
    \label{fig:EW-Genie_comp}
\end{figure}
The resulting flux-averaged 2p2h cross section is displayed in Fig.~\ref{fig:EW-Genie_comp}  as a function of the leading proton momentum \(p_N\).  The contributions of proton-proton  and proton-neutron  pairs in the final state are shown separately,  with the \(pp\)  channel clearly dominating across the relevant kinematic range.

To our knowledge the results of  Fig.~\ref{fig:EW-Genie_comp}  represents the first fully microscopic, relativistic, {\it semi-inclusive} computation of the 2p2h contribution to this class of processes. In the literature, the only available theoretical prediction for these data is the GENIE implementation~\cite{Dolan:2019bxf} of the SuSAv2-MEC model~\cite{RuizSimo:2016rtu,Megias:2016fjk}.  Our results display systematic differences with respect to GENIE, shown for comparison in  Fig.~\ref{fig:EW-Genie_comp} (hystogram): we consistently obtain higher strength, with the peak located at larger leading proton momenta.

To understand the origin of these discrepancies,  it should be kept in mind that the GENIE simulation relies on the {\it inclusive} 2p2h calculation of Ref.~\cite{RuizSimo:2016rtu} — based on the same theoretical framework as the present work, but limited to inclusive scattering — and it  requires certain assumptions in order to ``extract'' semi-inclusive predictions from the inclusive results. Such an extrapolation is,  in principle, not justified, as   semi-inclusive quantities depend on kinematic variables which are integrated out in the inclusive calculations. In contrast, our calculation deals correctly with all the kinematic variables, providing a more robust description. Therefore, differences between our model and the Monte Carlo outcomes are expected.
On the other hand, GENIE also includes additional nuclear effects absent from our calculation, such as final-state interactions (FSI), modeled via a semi-classical cascade. These interactions typically redistribute the strength of the cross section, shifting and broadening it toward lower final-proton energy and momentum. 
In particular, when considering only protons with momentum $p_N>500$ MeV, FSI tend to reduce the cross section. The cascade model accounts for multiple re-scattering processes of the ejected nucleon with other nucleons in the nucleus, which also influences the angular distribution of the final proton. In this context, kinematical angular cuts may further alter both the shape and strength of the cross section. 

A meaningful comparison could only be performed if  the present semi-inclusive calculation, which assumes that the two outgoing nucleons are described by plane waves, were implemented in Monte Carlo generators and propagated by consistent modeling of FSI. This  would constitute a major step forward in accurately describing neutrino-nucleus interactions in the 2p2h sector.

\section{Conclusion}

We have performed the extension of a microscopic calculation of the 2p2h  contribution from the electromagnetic $(e,e'p)$
 process\cite{Belocchi:2024rfp} to the weak sector, providing predictions for  the semi-inclusive  $(\nu_\mu,\mu p)$  cross section on carbon. 
 Unlike most of past literature, the model is able to provide observables as functions of both the leptonic and hadronic variables.
 
After validating the model  against inclusive results from the literature, we have presented semi-inclusive differential cross sections under realistic T2K-like kinematics.
Our results show that proton-proton  emission dominates over proton-neutron, with distinct energy dependencies. 
We have also analyzed in detail the roles of pion exchange, $\Delta$ excitation, and their interference, finding that the $\Delta$ resonance always dominates, while the interference is small but non-negligible and kinematics-dependent.

To connect with experimental observables, we have expressed results in terms of both detected and leading proton momentum, highlighting their impact on the cross section. Comparisons with GENIE simulations, based on inclusive inputs, revealed significant differences, emphasizing the need for more exclusive modeling in event generators.

While further developments are needed, for instance the inclusion of correlations and final state interactions, this work offers a practical and accurate tool to improve 2p2h  modeling in support of ongoing and future neutrino experiments.


\section*{Acknowledgements}
This work was partially supported by the Project NUCSYS funded by INFN; by the Grant BARM-RILO-24-01 funded by University of Turin; by the
“Planes Complementarios de I+D+i” program (Grant ASFAE/2022/022) by MICIU with funding from the
European Union NextGenerationEU and Generalitat Valenciana; by grant PID2023-147458NB-C21 funded by the European Union and by MCIN/AEI/10.13039/501100011033 .

\end{document}